\documentclass[aps,prl,reprint,floatfix]{revtex4-2}
\usepackage{amsmath}
\usepackage{graphicx}
\usepackage{capt-of}
\newenvironment{widefigure}
  {\par\onecolumngrid\vspace{10pt}}
  {\par\vspace{8.5pt}\twocolumngrid}
\graphicspath{{figures/}}
\usepackage{dcolumn}
\usepackage{xcolor}
\usepackage{siunitx}
\usepackage[section]{placeins}

\usepackage[hyperfootnotes=true]{hyperref}
\hypersetup{hidelinks}

\begin{document}
\raggedbottom

\title{Self-similar Worthington jets}

\author{Jos\'e M. Gordillo$^{1}$}
\email{jgordill@us.es}
\author{Javier Rodr\'iguez--Rodr\'{i}guez$^{2}$}
\author{Vatsal Sanjay$^{3}$}
\affiliation{$^1$\'Area de Mec\'anica de Fluidos, Departamento de Ingenier\'ia Aeroespacial y Mec\'anica de Fluidos, Universidad de Sevilla, Avenida de los Descubrimientos s/n 41092, Sevilla, Spain
}
\affiliation{$^2$ Department of Thermal and Fluids Engineering,
“Gregorio Millán Barbany” Institute, Universidad Carlos III de Madrid, Spain}
\affiliation{$^3$CoMPhy Lab, Department of Physics, Durham University, Science Laboratories, South Road, Durham
DH1 3LE, United Kingdom}

\date{\today}

\begin{abstract}

When a micron-sized bubble bursts, a conical cavity focuses into a singular Worthington jet. We derive self-similar Euler solutions whose exponent is selected by the cone angle. Simulations reveal a large Weber number diverging as the jet radius shrinks, contradicting the constant, order-unity value predicted by inertio-capillary theory with exponent $2/3$ in this regime. The theory remarkably improves predictions of experimental jet velocities. For water bubbles of radius $\SI{10}{\micro\meter}$, the predicted nanometric viscous cutoff suggests that non-continuum physics may regularize the singularity.

\end{abstract}

\maketitle

\begin{figure}
    \centering
    \includegraphics[width=0.8\linewidth]{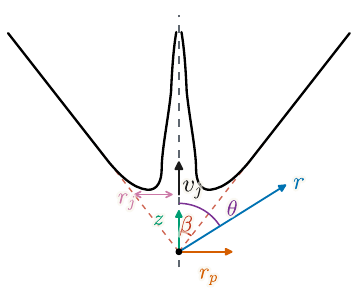}
    \caption{Schematic of the inertial conical-collapse problem. The gas cavity approaches a cone of semiangle $\beta$, described using the polar distance $r$ and angle $\theta$ measured from the apparent apex, or equivalently the axial and radial coordinates $z$ and $r_p$. The jet radius $r_j$ and velocity $v_j$ define the local observables used to test the self-similar scaling.}
    \label{fig:sketch_problem}
\end{figure}

\begin{figure*}[t]
\centering
\includegraphics[width=\textwidth]{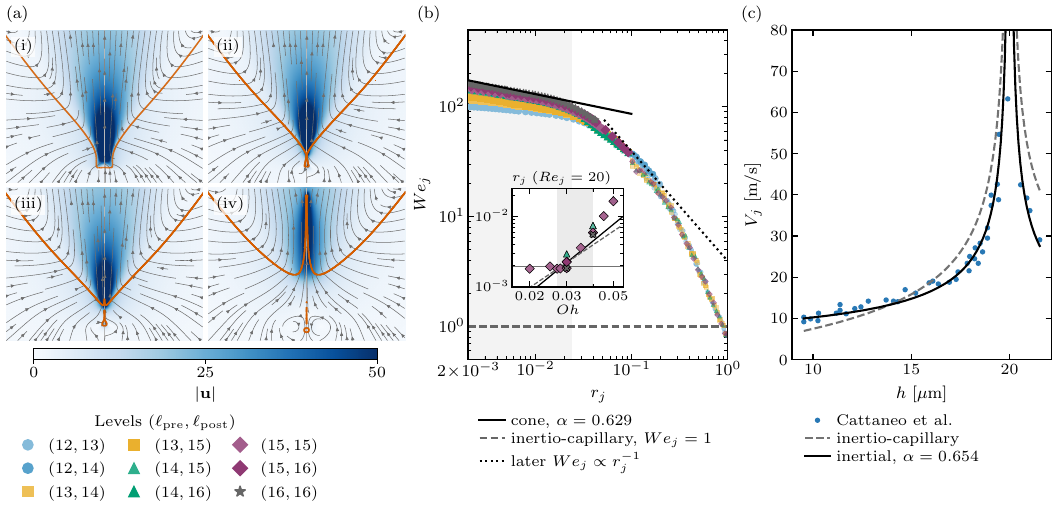}
\caption{Inertial focusing at $Bo=0$ and $Oh=0.03$. (a) Interface (orange), streamlines (grey), and velocity magnitude near jet inception. Snapshots (i)--(iv): $\tau'-\tau_0=-1.9\times10^{-3}$, $-1.3\times10^{-4}$, $+1.0\times10^{-4}$, and $+5.4\times10^{-3}$. (b) Local Weber number $We_j$ versus jet radius $r_j$. Symbols identify the refinement levels $(\ell_{\mathrm{pre}},\ell_{\mathrm{post}})$ before and after inception (see End Matter). Shading marks the conical-inertial window $0.002\le r_j\le0.024$. Solid, dashed, and dotted lines show our theory ($\alpha=0.629$), inertio-capillary scaling ($\alpha=2/3$), and the constant-flux scaling of Ref.~\cite{PRF2023}, respectively. Inset: radius at $Re_j=20$ versus $Oh$, with slope guides $\alpha/(2\alpha-1)$ for the same two exponents; the horizontal line marks the low-$Oh$ plateau. Grey shading in the inset marks the conical regime $0.027<Oh<0.04$. (c) Maximum jet velocities measured in Ref.~\cite{PRLCattaneo} (circles), compared with Eq.~(\ref{eq:experimental_jet_velocity}) (solid) and their inertio-capillary prediction (Eq.~(6) in Ref. \cite{PRLCattaneo}, dashed).}
\label{fig:data-figs}
\end{figure*}

Bubble bursting ejects Worthington jets whose breakup transfers material across liquid--gas interfaces,
from sea spray \cite{MacIntyre,Bigg,PNAS,AnnRevDeike,PNASWang,Abyss} and rain-induced aerosols \cite{RainPorous} to sparkling wines
\cite{SeonLigerBelair2017} and electrochemical sprays \cite{Bashkatov2025}.
The jet dynamics and minimum drop sizes are set by capillary-wave focusing \cite{Blake1993,Duchemin,Walls2015,krishnan2017,PRFBird,Deike,Berny}. The radii $R$ of the bubbles for which submicrometric droplets can be produced via bubble bursting jets at a surfactant-free or negligibly contaminated interface of a liquid of density $\rho$, dynamic viscosity $\mu$ and interfacial tension coefficient $\sigma$, are characterized by a value of the Ohnesorge number $Oh=\mu/\sqrt{\rho R\sigma}\approx0.03$ \cite{Walls2015,PRFBird}; for water, this means micron-sized bubbles with $Bo=\rho gR^2/\sigma\ll1$, with $g$ indicating gravity \footnote{In the following, lower-case letters indicate dimensionless variables defined using $R$, the capillary velocity $V_c=\sqrt{\sigma/(\rho R)}$, and $\rho V_c^2$ as the characteristic values of length, velocity and pressure. In the limit $Bo\rightarrow0$, the pre-inception bubble-bursting dynamics are controlled by $Oh$, or equivalently $La=Oh^{-2}$; gas properties become relevant after ejection through the drag exerted on the jet tip and emitted droplets \cite{JFM2020,Tian2023}.}. In this regime, capillary waves deform the bubble into a conical cavity of semiangle $\beta$ and entrap a tiny bubble beneath the cavity before the fastest, thinnest jet is emitted \cite{JFM2019} (Figs.~\ref{fig:sketch_problem} and~\ref{fig:data-figs}(a)(i)--(iv)). The local question is what similarity law governs the focusing flow after this conical cavity has formed.

We denote raw dimensionless simulation time by $\tau'=tV_c/R$ and time relative to the singularity by $\tau=\tau'-\tau_0$, where the singularity time $\tau_0$ depends on $Oh$. In the classical inertio-capillary interpretation, dynamic and capillary pressures balance during collapse, giving $r_j\propto|\tau|^{2/3}$ and a constant local Weber number $We_j=v_j^2r_j=\mathcal{O}(1)$ \cite{Zeff,Duchemin,Sierou,PRLEggers,PRLCattaneo}. Interface collapse alone does not probe this balance as both $2/3$ and a nearby exponent can collapse the evolving shapes \cite{Zeff,PRLEggers,PRLCattaneo}. Our direct numerical simulations instead give $We_j=\mathcal{O}(10^2)$, diverging as $r_j$ decreases (Fig.~\ref{fig:data-figs}(b)). These results establish a physical inconsistency in applying inertio-capillary similarity to the singular Worthington jets considered here \cite{Ghabache,PRLGanan,PRFBird,Deike,PRLEggers,JFM2020,JFMGanan,Gananfinespray,PRLCattaneo,Eshima2026,JFM2026}: the assumed constant, order-unity Weber number contradicts the large, diverging value at the jet base. The underlying similarity solutions \cite{Zeff,Sierou} remain applicable to inertial ligament pinch-off but the force balance of the singular Worthington jet is different.

\begin{figure}
\centering
\includegraphics[width=\columnwidth]{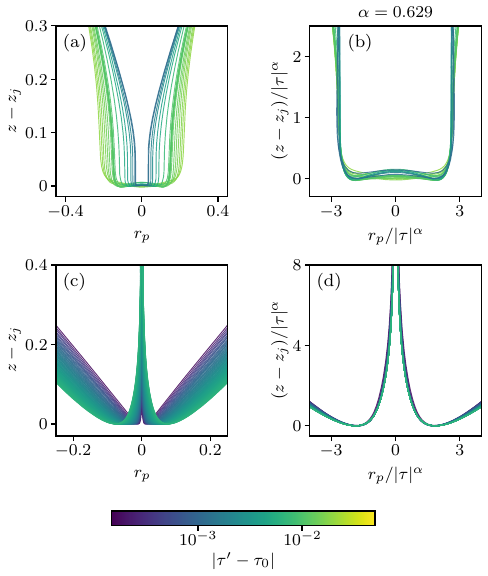}
\caption{Self-similar interface evolution at $Bo=0$, $Oh=0.03$. The upper and lower rows show cavity collapse before inception and the emitted jet, respectively. Panels (a,c) show unscaled interfaces, shifted by the instantaneous axial reference $z_j$; (b,d) show the same profiles scaled with $|\tau|^\alpha$, where $\alpha=0.629$. The colour scale uses $|\tau'-\tau_0|$ with $\tau_0=0.494430$. The complete comparison with $\alpha=2/3$ appears in Supplemental Fig.~S3.}
\label{fig:collapse}
\end{figure}

We resolve this inconsistency by deriving self-similar Euler solutions of the second kind, with the cone semiangle $\beta$ selecting the value of the exponent $\alpha(\beta)\lesssim 2/3$ in $r_j\propto \tau^\alpha$ for the measured bubble-bursting cone angles, finding that capillary effects are negligible. Indeed, for $Oh=0.03$, the measured $\beta=38.4^\circ$ gives $\alpha=0.629$ and, hence, our theory predicts, in agreement with the numerical results, the value of the exponent $\gamma<0$ in $We_j\propto r_j^{\gamma}$, which implies $We_j\to\infty$ as $r_j\to0$ (Fig.~\ref{fig:data-figs}(b)). Moreover, the rescaled numerical interfaces collapse onto a universal shape before and after jet inception (Fig.~\ref{fig:collapse}). Our geometry-selected theory also gives remarkably better agreement with the experimental jet velocities in an acoustically driven conical cavity than the inertio-capillary expression (Fig.~\ref{fig:data-figs}(c)) \cite{PRLCattaneo}. The jet dynamics determine the sizes and velocities of bubble-bursting drops that feed sea-spray aerosols \cite{PNAS} and the thicknesses and velocities of jets produced by the acoustic forcing of microbubbles, with applications in targeted drug delivery and biofilm removal \cite{NatPhysCatt,PRLCattaneo}.
The geometry-selected theory describes local inertial collapse with an approximately equipotential conical boundary, independently of how the cavity forms.

Next, we solve the conical collapse of the cavity as a local inertial-focusing problem. Capillarity and viscosity, through $Oh$, select whether the system enters the singular regime --- when the tiny bubble is entrapped --- and set the cone angle $\beta(Oh)$ and jet inception time in the small-$Bo$ regime. More generally, the cavity evolution before inception sets $\beta(Oh,Bo)$ \footnote{The Bond number changes the initial cavity shape, while $Oh$ controls capillary-wave damping.}. Once the similarity window $r_j\ll r\ll1$ exists with $r$ the polar distance, see Fig.~\ref{fig:sketch_problem}, the dominant balance must be inertial owing to the very large values of the local Weber number. The exponent in $r_j\propto \tau^{\alpha(\beta)}$ is therefore not universal but fixed by the far-field cone geometry through $\alpha(\beta)=1/(2-\nu(\beta))$, with $\nu(\beta)$ fixed by the condition $P_\nu(-\cos\beta)=0$, where $P_\nu$ indicates the Legendre function of degree $\nu$. Measuring the incoming angle fixes the exponent independently of a fit to the jet dynamics. Dependence on the initial or far-field cavity geometry, also observed in recent shallow-layer bubble-bursting jets \cite{YangShallow2026}, is a signature of a second-kind similarity rather than a secondary correction.

The experimental measurements in \cite{PRLCattaneo} test the velocity prediction; the measured growth of $We_j$ as $r_j \to 0$ tests the force balance, and the trend $\alpha[\beta(Oh)]<2/3$ tests the exponent (see End Matter). For water bubbles of radius $R\sim\SI{10}{\micro\meter}$, the predicted viscous cutoff lies at nanometric scales, where additional physics must enter before viscosity can regularize the continuum-incompressible singularity.

\emph{Conical similarity.} --- When the production of vorticity at the interface can be neglected, the velocity field $\mathbf{v}$ can be expressed as the gradient of a velocity potential $\phi$, $\mathbf{v}=\nabla\phi$. In the incompressible limit, $\phi$ verifies the Laplace equation $\nabla^2\phi=0$, which must be solved subject to dynamic and kinematic boundary conditions at the interface $F=r_p-r_s(z,\tau)=0$, with $r_s$ the radial interface position in cylindrical coordinates $(z,r_p)$. The spherical coordinates $(r,\theta)$ are centred at the apparent apex of the cone (see Fig.~\ref{fig:sketch_problem}). In the limit in which capillary and viscous stresses are negligible, the dynamic boundary condition is the Euler--Bernoulli equation particularized at $F=0$, $\partial\phi/\partial \tau+|\nabla\phi|^2/2=0$, whereas the kinematic boundary condition reads $\partial F/\partial \tau+\nabla\phi\cdot\nabla F=0$.

Far from the jet base in the local similarity limit, $r/r_j\rightarrow\infty$ with $r\ll1$, the interface converges to a cone of semiangle $\theta=\beta$, so the local problem lacks a spatial scale. Thus, it is natural to seek for a local, self-similar solution. To investigate the structure of the solution, we make use of the scaled variables proposed in Ref.~\cite{Zeff},
\begin{equation}
\begin{split}
&\phi=\tau^\delta\,h\left[\frac{z'-\ell_0(Oh)}{\tau^\varepsilon},\frac{r_p}{\tau^\varepsilon}\right] \\ &\mathrm{and}\quad
r_s(z,\tau)=\tau^\varepsilon\,g\left[\frac{z'-\ell_0(Oh)}{\tau^\varepsilon}\right]\, \label{ansatz}
\end{split}
\end{equation}
with $\tau>0$ for the post-inception solution, and $z=z'-\ell_0(Oh)$ and $r_p$ the axial and radial cylindrical coordinates (see Fig.~\ref{fig:sketch_problem}). Before inception, $\tau<0$ and interface lengths are rescaled with $|\tau|^\varepsilon$. The system composed of the Laplace equation, the Euler--Bernoulli equation, and the kinematic boundary condition can be written in terms of the variables in Eq.~(\ref{ansatz}) if $\delta=2\varepsilon-1$ \cite{Zeff}, with this condition arising from the balance between the temporal and the convective derivatives in the dynamic and kinematic boundary conditions. So far, equation $\delta=2\varepsilon-1$ constitutes the only restriction for self-similar solutions to exist. However, an additional condition that the exponents $\delta$ and $\varepsilon$ must fulfill is deduced by imposing the behaviour of the velocity potential $\phi$ in the far field. The constant-flux condition in \cite{PRF2023} describes conical collapse, including slender cones ($\beta\ll1$), and bubble-bursting jets emitted from truncated cones at $Oh\lesssim0.02$. Truncation removes the apex without fixing the conical sidewall angle $\beta$. The far-field velocity potential then satisfies

\begin{equation}
r_p\frac{\partial\phi}{\partial r_p}\rightarrow -q_\infty\quad\mathrm{for}\quad r_p/r_j\rightarrow\infty\, ,\label{qinfty}
\end{equation}
with $q_\infty$ the value of the time-independent far-field flow rate per unit length along the axis of the cone, fixed at the instant when the jet starts to be issued. Notice that Eqs.~(\ref{ansatz}) and (\ref{qinfty}) imply that $\delta=0$ and, hence, $\varepsilon=1/2$ and $r_j(\tau)\propto \tau^{1/2}$, with $We_j\propto \tau^{-1/2}\propto r^{-1}_j$, in agreement with numerical results \cite{PRF2023} --- see also Fig.~\ref{fig:data-figs}(b).

However, for $0.027 \lesssim Oh \lesssim 0.04$, jets are issued right after a tiny bubble is entrapped at the bottom of the cavity as a consequence of the collapse of a conical cavity of semiangle $\beta=\mathcal{O}(1)$, for which the far-field boundary condition in Eq.~(\ref{qinfty}) is no longer valid. In this case, which corresponds to the thinnest and fastest jets, the early dynamics admit an intermediate spatial region $r_j(\tau)\ll r\ll1$ for $\tau\ll1$ (see Fig.~\ref{fig:sketch_problem}). The velocity decays in the local similarity far field, where the quadratic velocity term is negligible to leading order \cite{wagner1932}. The Euler--Bernoulli condition on $F=0$ then gives $\partial\phi/\partial\tau=-|\nabla\phi|^2/2\simeq0$. The material derivative satisfies $D\phi/D\tau=+|\nabla\phi|^2/2\simeq0$ under the same approximation. Thus an initially equipotential interface, with its potential set to $\phi(\tau=0)=0$, remains approximately equipotential in this short-time limit, giving $\phi(\theta=\beta)\simeq0$ in the intermediate region. The results in Figs. 1(b) and 6 of \cite{JFMGanan} and the literature results support this condition; the velocity field in Fig.~\ref{fig:data-figs}(a)(ii,iii) is normal to the conical surface, showing that this surface is approximately equipotential in the intermediate region. Hence, for $0.027\lesssim Oh\lesssim 0.04$, the far field boundary condition driving the ejection of the jet must be a solution of the Laplace equation satisfying the condition $\phi(\theta=\beta)=0$, which implies that

\begin{equation}
\phi\sim A r^\nu P_\nu(-\cos\theta)=A \tau^{\varepsilon \nu}\left(r/\tau^\varepsilon\right)^\nu P_\nu(-\cos\theta) \label{farfieldBC}
\end{equation}
in the similarity far field $r_j\ll r\ll1$, with the time-independent constant $A$ fixed by the velocity in the far field at the instant when the jet starts to be issued. Equation (\ref{farfieldBC}), together with the far-field condition $\phi(\theta = \beta) = 0$, imposes that the degree of the Legendre function $\nu$ must satisfy the condition

\begin{equation}
P_\nu(-\cos\beta)=0\, .\label{condition}
\end{equation}

Fig.~\ref{fig:nu_vs_beta} in the End Matter
shows the solution of Eq.~(\ref{condition}) and, thus, the value of the degree $\nu$ for different values of the cone semiangle $\beta$. Matching the potential in Eq.~(\ref{ansatz}) to Eq.~(\ref{farfieldBC}) implies that self-similar solutions will exist if $\delta=\nu \varepsilon$ and $\delta=2\varepsilon-1$, namely, if $\varepsilon=1/(2-\nu(\beta))$. Hence, for the present family of self-similar solutions, $r_j(\tau)\propto \tau^{1/(2-\nu(\beta))}$, $v_j(\tau)\propto \tau^{(\nu(\beta)-1)/(2-\nu(\beta))}$ and $We_j(\tau)\propto \tau^{(2\nu(\beta)-1)/(2-\nu(\beta))}$. In view of the values of $\nu(\beta)$ given in Fig.~\ref{fig:nu_vs_beta} and for the measured bubble-entrapment range $35^\circ\lesssim\beta\lesssim41^\circ$ ($\beta=38.4^\circ$ at $Oh=0.03$; Fig.~\ref{fig:nu_vs_beta}), $1/(2-\nu(\beta))\lesssim 2/3$ and $We_j\gg 1$ for $\tau\ll 1$, which is a necessary condition for fast and slender Worthington jets to be issued after the collapse of a cavity \cite{JFM2019}.

The values of $\nu(\beta)$ in Fig.~\ref{fig:nu_vs_beta} (End Matter) show that $\nu(\beta)\rightarrow 0$ and $\alpha(\beta)\to1/2$ as $\beta\rightarrow0$, consistent with the far-field condition in Eq.~(\ref{qinfty}), used in \cite{PRF2023} to predict, up to prefactors, the ejection of Worthington jets following the collapse of cones with $\beta\ll 1$ or slender axisymmetric cavities of arbitrary shape, like parabolas. These theoretical predictions are in excellent agreement with experiments and numerical simulations \cite{PRL2024}. Also, $r_j(\tau)\propto \tau^{2/3}$ when the semiangle $\beta$ coincides with the one characterizing Taylor's cone in electrosprays, $\beta=49.3^\circ$ \cite{Taylor}. Here $\nu=1/2$, so $\alpha=2/3$ is a special case of the present inertial family. Its $We_j$ is constant, without requiring its magnitude to be of order unity. The exponent $2/3$ alone therefore cannot establish an inertio-capillary balance. Here, we stress that one of the main advances over \cite{PRF2023} is that our theory predicts the near-inception jet dynamics observed in simulations and experiments for conical cavities with $\beta=\mathcal{O}(1)$, with the equipotential condition $\phi(\theta=\beta)=0$ selecting $\nu(\beta)$ and hence $\alpha(\beta)=1/(2-\nu(\beta))$, whereas constant-flux forcing gives $r_j\propto\tau^{1/2}$ with $\beta$-dependent prefactors. In fact, it is shown below that our present theory describes the conical-collapse regime relevant to sea spray \cite{AnnRevDeike} and to acoustically driven jets used in drug delivery and biofilm removal \cite{NatPhysCatt,PRLCattaneo}.

\emph{DNS and experimental validation.} --- We express three local quantities as functions of the jet radius $r_j(\tau)$, namely the flow rate per unit length along the axis of the cone $q_j=v_j r_j$, the flow rate feeding the jet $Q_j=\pi r_j q_j$, and the local Weber number $We_j=v^2_j r_j=q^2_j/r_j$. After some algebra, we find $q_j(\tau)\propto r_j^{(2\alpha(\beta)-1)/\alpha(\beta)}$, $Q_j\propto r_j^{(3\alpha(\beta)-1)/\alpha(\beta)}$, $We_j\propto r_j^{(3\alpha(\beta)-2)/\alpha(\beta)}$, with $\alpha(\beta)=1/(2-\nu(\beta))$ and $\nu(\beta)$ fixed by the condition $P_\nu(-\cos\beta)=0$. The results depicted in Fig.~\ref{fig:data-figs}(b) and Fig.~\ref{fig:oh-levels}(a) in the End Matter compare our theoretical results above with the numerical values of $We_j(r_j)$ and $Q_j(r_j)$ calculated using \texttt{Basilisk C} \cite{Popinet2009,Popinet2015,Vatsal2021} for $Oh=0.03$, with $\beta$ determined from the numerical simulations, showing that inertia governs the jet-base dynamics, as predicted by our theory. The plot corresponding to $We_j(r_j)=q^2_j/r_j$ in Fig.~\ref{fig:data-figs}(b) (see also Fig.~\ref{fig:oh-levels}(b) in the End Matter) reveals that our theory clearly departs from the inertio-capillary balance $We_j=\mathrm{const}=\mathcal{O}(1)$ and captures the observed growth of $We_j$ toward the singularity ($We_j=\mathcal{O}(10^2)$ over $0.002\le r_j\le0.024$); notice also that $We_j\propto r^{-1}_j$ for larger times after jet inception \footnote{The spatial region in which streamlines are horizontal in Fig.~4(b) of Ref.~\cite{PRF2023} corresponds to the far-field boundary condition in Eq.~(\ref{qinfty}): when, at later times after jet inception the vertical position of the base of the jet reaches this region, $r_j\propto \tau^{1/2}$, $We_j\propto \tau^{-1/2}\propto r^{-1}_j$ for all values of $Oh$, no matter the value of $\beta$ \cite{PRF2023}}.
Moreover, Fig.~\ref{fig:collapse} (see also Supplemental Fig.~S3 and Supplementary Videos~1 and~2) shows the collapse of the rescaled shapes before and after inception, strongly supporting the predicted self-similar structure. The same profiles also collapse with $\alpha=2/3$ (Supplemental Fig.~S3). The distinction between the two theories is therefore the physical balance revealed by $We_j$. Notice also that the reason why it has been long thought \cite{Zeff,PRLEggers,PRLCattaneo} that the collapse of clean capillary cavities is driven by a balance between dynamic and capillary pressures relies on the small differences between the inertio-capillary exponent $\alpha=2/3$ and the correct result $\alpha(\beta\approx 40^\circ)\approx0.63$.

Next, we compare in Fig.~\ref{fig:data-figs}(c) our theory with the experimental values of the maximum jet velocity reported in Fig.~5(a) of Ref.~\cite{PRLCattaneo}, corresponding to a collapsing conical cavity formed by converging, acoustically driven surface waves.
To carry out the comparison, we need to define a length scale in their flow equivalent to our initial bubble radius, $R$, which we choose to be their critical surface wave amplitude $h_c \approx \SI{20}{\micro\meter}$. Using our previous result  $q_j=v_j r_j\propto r_j^{(2\alpha(\beta)-1)/\alpha(\beta)}$ and defining the dimensionless smallest length scale of the jet as $r_m = |h - h_c|/h_c$, where $h$ is their wave amplitude, we can write
\begin{equation}
    V_j = C_M \, \frac{\sigma}{\mu} \, Oh_c^{-2(\alpha-1)/\alpha} \, r_m^{(\alpha-1)/\alpha},
    \label{eq:experimental_jet_velocity}
\end{equation}
where $Oh_c = \mu/\sqrt{\rho \sigma h_c} \approx 0.0264$ and $C_M$ is a numerical prefactor of order unity.
The experimental jet velocity of Fig.~5(a) of Ref.~\cite{PRLCattaneo} is compared in Fig.~\ref{fig:data-figs}(c) with the prediction of our equation (\ref{eq:experimental_jet_velocity}), using $C_M = 4.6$ and $\alpha = 0.654$, corresponding to $\beta = 46^\circ$ (as measured from Fig.~3(a) in Ref.~\cite{PRLCattaneo}). We can see that the agreement with the experiments is visibly better than the one obtained using the inertio-capillary theory (Eq.~(6) in Ref.~\cite{PRLCattaneo}), proving that our theory can be used to achieve better quantitative predictions of existing experimental data.

\emph{Viscous cutoff.} --- Viscosity becomes comparable to inertia when $Re_j=v_jr_j/Oh\sim1$ \cite{JFM2019}. Combining this condition with $v_jr_j\propto r_j^{(2\alpha-1)/\alpha}$ gives $r_m\propto Oh^{\alpha/(2\alpha-1)}$ and $R_m=Rr_m\propto[\mu^2/(\rho\sigma)]Oh^{(2-3\alpha)/(2\alpha-1)}$. In the simulations, $Re_j$ remains above unity. We choose $Re_j=20$ and measure the corresponding radius, for which the two finest grids agree (inset of Fig.~\ref{fig:data-figs}(b) and Fig.~\ref{fig:oh-levels}(c) in the End Matter). This gives a grid-independent diagnostic of the regularization of the resolved branch \footnote{Supplemental Fig.~S2 tests the sensitivity of the radii in the Fig.~\ref{fig:data-figs}(b) inset and Fig.~\ref{fig:oh-levels}(c) to the chosen Reynolds-number threshold. The increase with $Oh$ and the agreement across different grid resolutions persist at $Re_j=25$ and 30.}. The measured radius is an upper-bound proxy for $r_m$ as the $Re_j=1$ crossing remains at the mesh floor. Due to the fact that $\alpha(\beta)<2/3$ \footnote{See \cite{Thoroddsen_18,Thoroddsen_2020,Tian2023}, where experimental exponents $\alpha<2/3$ are also reported.}, the minimum jet radius and the corresponding maximum achievable velocity are respectively smaller (radius) and larger (velocity) than the ones deduced using $\alpha=2/3$, which gives $R_m\sim\SI{100}{\nano\meter}$ \cite{JFM2020,AnnRevDeike,VillermauxNexus}. Indeed, for $Oh=0.03$, the cone fit \cite{BurstingBubbleCode} gives $\beta=38.4^\circ$, $\alpha\simeq0.63$, $R_m\sim 0.2 \mu^2/(\rho\sigma)$, and the maximum velocity deduced using the definition of $Re_j$ is $V_M \approx 4 \,\sigma/\mu$; using water properties, $R_m\sim\SI{3}{\nano\meter}$ and $V_M\sim\SI{320}{\meter\per\second}$: these values indicate that viscosity is unable to regularize the singularity within the range of validity of the continuum description, and also suggest that, thanks to their purely inertial nature, \emph{if continuum axisymmetry was preserved down to the instant of jet inception} \cite{JFM2010}, bubble bursting jets in clean liquids could produce a flux of aerosol particles with sizes ranging from nanometres --- not hundreds of nanometres, as it was previously thought \cite{JFM2020,AnnRevDeike,VillermauxNexus} --- to microns.

\emph{Conclusion.} --- We have derived a family of self-similar solutions of the incompressible Euler equations for the collapse of a conical gas cavity of semiangle $\beta$. The jet radius and velocity scale as $r_j(\tau)\propto \tau^{1/(2-\nu(\beta))}$ and $v_j(\tau)\propto \tau^{(\nu(\beta)-1)/(2-\nu(\beta))}$, with $\nu(\beta)$ the degree of the Legendre function satisfying $P_\nu(-\cos\beta)=0$. For the measured bubble-entrapment range $35^\circ\lesssim\beta\lesssim41^\circ$, this gives $0.62\lesssim1/(2-\nu)\lesssim0.64$, close enough to $2/3$ to explain why inertio-capillary scalings appeared plausible.
The evidence for the validity of our result comes directly from $We_j$, not only from the fitted exponent. For the measured bubble-bursting cone angles, $We_j(\tau)\propto \tau^{(2\nu-1)/(2-\nu)}$ diverges as $\tau\rightarrow0$, and the DNS gives both $We_j\gg1$ and collapse of the free surface near the jet base over the pre- and post-inception windows in Fig.~\ref{fig:collapse} when lengths are scaled with $|\tau|^{1/(2-\nu[\beta(Oh)])}$. For these small-$Bo$ cases, $Oh$ controls access to the singularity by selecting $\beta$, the inception time, and the regularization scale; once the similarity window exists, the local collapse is governed by inertia. For water, we find that viscosity is unable to regularize the divergent self-similar solution within the continuum approach.
The predicted nanometric viscous cutoff calls for a description of how inertial focusing is regularized in a real liquid. Future work could match the present similarity solution to a molecular description of jet inception, while three-dimensional simulations would be needed to test whether azimuthal perturbations interrupt focusing at larger scales. Following the resulting jet through breakup, including thermal fluctuations \cite{MoselerLandman2000,Eggers2002}, would establish whether nanometric jet radii lead to nanometric droplets, and how the geometry-selected collapse shapes the smallest sea-spray aerosols.
The range of validity of this inertial description could also be tested across other routes to singular jet formation, including Faraday-wave collapse \cite{Zeff}, air-cavity collapse during drop recoil on superhydrophobic solids \cite{ZhangForces2022}, and the singular-jet regime after drop impact on a deep liquid pool \cite{Thoroddsen_18,Thoroddsen_2020,Tian2023,Michon}. Comparisons from finite cone angles to the slender limit could identify where the approximately equipotential conical boundary or inertial balance fails.

\begin{acknowledgements}
This research work has been partially supported by the Grants PID2024-156545NB-I00 and PID2023-146809OB-I00, financed by the Spanish MICIU/AEI/10.13039/501100011033 and by ERDF/EU. V.S. acknowledges start-up funding from Durham University. This work made use of the Hamilton HPC Service of Durham University. The simulation work was also carried out on the national e-infrastructure of SURFsara, a subsidiary of SURF cooperation, the collaborative ICT organisation for Dutch education and research; this simulation work was sponsored by NWO~--- Domain Science for the use of supercomputer facilities.
\end{acknowledgements}

\noindent Code availability. The solver, case files, and post-processing scripts required to reproduce the figures in this Letter are available at the public CoMPhy Lab repository \cite{BurstingBubbleCode}.

\bibliography{Gordillo_bibv2}

\section*{\texorpdfstring{\NoCaseChange{End Matter}}{End Matter}}

\begin{widefigure}
\noindent\begin{minipage}{\textwidth}
    \centering
    \includegraphics[width=\textwidth]{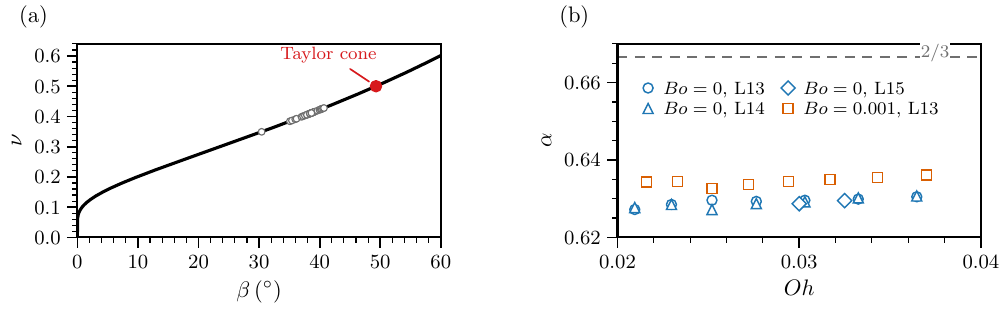}
    \captionof{figure}{(a) Values of $\nu$ solving Eq.~(\ref{condition}) as a function of the cone angle $\beta$; the grey symbols are the cone angles measured in the DNS. The red dot marks Taylor's cone, $\beta=49.3^\circ$, $\nu=0.5$. (b) Exponent $\alpha=1/(2-\nu)$ as a function of $Oh$ for $Bo=0$ (levels 13--15) and $Bo=10^{-3}$ (level 13), using the same cone fits; marker shapes distinguish refinement levels. The dashed line marks $\alpha=2/3$.
    }
    \label{fig:nu_vs_beta}
    \end{minipage}
\end{widefigure}

\subsection*{\texorpdfstring{\NoCaseChange{Numerical methods and data availability}}{Numerical methods and data availability}}

We solve the axisymmetric two-phase incompressible Navier--Stokes equations with the opensource free-software language \textit{Basilisk C} \cite{Popinet2009,Popinet2015,Vatsal2021}. The interface is represented using a geometric volume-of-fluid field $c$, with density and viscosity interpolated in the one-fluid formulation. At $Bo=0$, the dimensional governing equations are
\begin{align}
\nabla\cdot\mathbf{u}
&\;= 0,
\label{eq:dns-incompressibility}\\
\rho(c)\left(\partial_t\mathbf{u}+\mathbf{u}\cdot\nabla\mathbf{u}\right)
&\;= -\nabla p+\nabla\cdot\left[2\mu(c)\mathbf{D}\right]+\sigma\kappa\delta_s\mathbf{n},
\label{eq:dns-momentum}
\end{align}
where $\mathbf{D}=\left[\nabla\mathbf{u}+(\nabla\mathbf{u})^T\right]/2$. Lengths, velocities, time, and pressure are scaled with $R$, $V_c=\sqrt{\sigma/(\rho_l R)}$, $R/V_c$, and $\sigma/R$, respectively. We fix $\rho_g/\rho_l=10^{-3}$ and $\mu_g/\mu_l=2\times10^{-2}$; all results reported in the Letter use $Bo=0$ and $Oh\simeq0.03$, unless stated otherwise.
The additional $Oh=0.028$ and $0.04$ series are shown in Fig.~\ref{fig:oh-levels}.

For $Bo=0$, the initial condition is a spherical bubble of radius $R$ intersecting a flat free surface. The point contact is regularized by a capillary bridge of size $\delta_b=0.01R$, leaving an initial opening of radius $\simeq2\delta_b$. The computational domain spans $-6\le z'\le4$ and $0\le r_p\le10$ for the production cases; the lower axial boundary is a no-slip, no-penetration wall, the upper axial boundary is an outflow, and $r_p=0$ is the axis of symmetry.

Adaptive mesh refinement is applied to the volume fraction, velocity, and curvature fields. To resolve the focusing singularity at jet inception, we developed a feature-tracking refinement criterion that follows the largest incoming curvature wave before inception and the newly formed jet thereafter. The interface shapes in Fig.~\ref{fig:collapse} use $MAXlevel=15$, corresponding to a finest cell size $\Delta/R=10/2^{15}\simeq3.1\times10^{-4}$, while the flux comparison in Fig.~\ref{fig:data-figs} spans post-inception levels 13--16.

Grid sensitivity is assessed in Fig.~\ref{fig:data-figs} using pairs $(\ell_{\mathrm{pre}},\ell_{\mathrm{post}})$. Here $\ell_{\mathrm{pre}}$ is the temporary ceiling while the focusing curvature wave is tracked before inception; once the jet forms, the mesh is released to the post-inception ceiling $\ell_{\mathrm{post}}$.

\subsection*{\texorpdfstring{Solution to the equation $P_\nu(-\cos\beta) = 0$}{Solution to the equation Pnu(-cos beta) = 0}}

For each $\beta$, Eq.~(\ref{condition}) is solved as a scalar root-finding problem for the real degree $\nu$ of $P_\nu(-\cos\beta)$, choosing the branch continuous with Taylor's cone, $\beta=49.3^\circ$ and $\nu=1/2$. Fig.~\ref{fig:nu_vs_beta}(a) shows this mapping; panel (b) converts the DNS-measured cone angles into $\alpha=1/(2-\nu)$ as a function of $Oh$. The exponents are systematically below $2/3$ and vary with $Oh$, providing additional evidence against a fixed inertio-capillary exponent and in favour of the inertial conical-collapse similarity.

\begin{widefigure}
    \noindent\begin{minipage}{\textwidth}
            \centering
            \includegraphics[width=\textwidth]{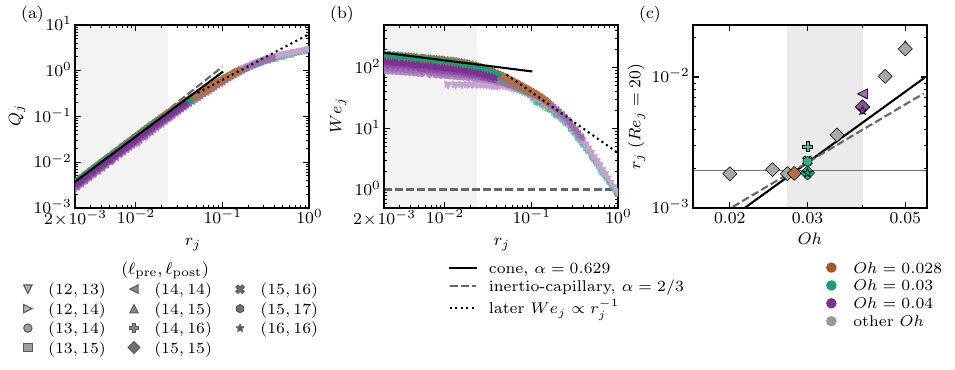}
            \captionof{figure}{Dependence on $Oh$ and refinement. (a) Jet-feeding flux $Q_j$ versus $r_j$ for $Oh=0.028$, $0.03$ and $0.04$. (b) Corresponding Weber-number trajectories, with $We_j=Q_j^2/(\pi^2r_j^3)$. In (a,b), colour denotes $Oh$, darker shades indicate higher focusing levels $\ell_{\rm pre}$, and symbols identify the pre- and post-inception refinement pairs. Shading marks the conical-inertial regime. The guides in (b) are those of Fig.~\ref{fig:data-figs}(b), with the dashed line marking $We_j=1$. (c) Radius at fixed $Re_j=20$, showing the full dataset and the same slope guides as in the inset of Fig.~\ref{fig:data-figs}(b). Grey points denote the additional $Oh$ values. In (c), grey shading marks the conical regime $0.027<Oh<0.04$.}
        \label{fig:oh-levels}
        \end{minipage}
    \end{widefigure}

\end{document}


\maketitle

\section*{Cone-angle extraction and sensitivity}

\begin{figure}[!ht]
\centering
\includegraphics[width=\textwidth]{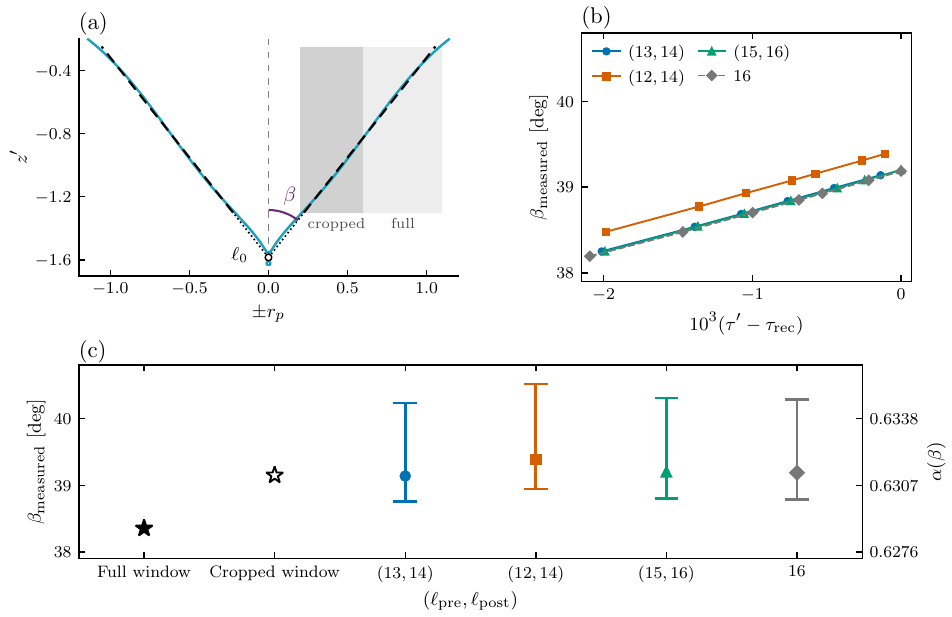}
\caption{Cone-angle extraction and sensitivity. (a) Cavity interface at $Oh=0.03$, with the full fitting window in light grey and its cropped part in darker grey. The full-window fit (dashed line) gives $\beta=38.4^\circ$; dotted extensions reach the apparent apex $\ell_0$, and the arc measures $\beta$ from the axis. (b) Evolution of $\beta$ before reconnection, using the cropped window at the available snapshot times. (c) Angles at the last pre-reconnection snapshot of each simulation, labelled by pre/post refinement levels, with ranges from nine fitting windows. The right axis gives the corresponding $\alpha(\beta)$. Filled and open stars mark the full and cropped fits to the same interface in (a).}
\label{fig:cone-sensitivity}
\end{figure}

At the snapshot nearest to jet inception, we fit the cavity flank to $r_p=mz'+c$ by unweighted least squares on the interface-facet endpoints within $-1.30\le z'\le-0.25$ and $0.20\le r_p\le1.10$ (Fig.~\ref{fig:cone-sensitivity}(a)). The fitted line gives the semiangle from the axis, $\beta=\arctan m$, and the apparent apex, $\ell_0=-c/m$. This full-window fit gives $\beta=38.356^\circ$. Solving $P_\nu(-\cos\beta)=0$ then determines $\alpha=1/(2-\nu)$ independently of the jet-flux fit.

To compare with the other interface sequences, which extend only to $r_p<0.6$, we crop the radial fitting window to $0.20\le r_p\le0.60$ and retain the same axial limits. For the same interface in Fig.~\ref{fig:cone-sensitivity}(a), cropping alone increases $\beta$ to $39.151^\circ$, a shift of $0.795^\circ$. Within the cropped window, $\beta$ rises by $0.89$--$0.99^\circ$ during the last $\tau \approx 2\times10^{-3}$ before reconnection (Fig.~\ref{fig:cone-sensitivity}(b)). Across the nine windows defined by $r_{p,\min}\in\{0.15,0.20,0.25\}$ and $r_{p,\max}\in\{0.50,0.55,0.60\}$, the fitted angle varies by $1.48$--$1.57^\circ$ at the last pre-reconnection snapshot of each simulation (panel c).

The solver, cone-fitting routine and paper-processing scripts are available in the \href{https://github.com/comphy-lab/Bursting-Bubble}{public CoMPhy Lab repository}.

\section*{Flux exponents and refinement}

At $Oh=0.03$, the jet has $We_j\simeq80$--$142$ and $Re_j\simeq23$--$51$ over the reference interval $0.005\le r_j\le0.023952$. The Weber number is large and increases as $r_j$ decreases; together with $Re_j>20$, this supports the inertial balance within this window. We take the theoretical exponent $0.629$ from the cone geometry and compare it with the numerical flux slopes measured independently over finite radius intervals.

For a flux law $Q_j\propto r_j^s$, the corresponding Weber-number law is $We_j\propto r_j^{2s-3}$ and the effective exponent is $\alpha_{\rm eff}=1/(3-s)$. The two slopes are therefore algebraically equivalent diagnostics. We fit the plotted jet-base flux $Q_j$ separately for each refinement history $(\ell_{\rm pre},\ell_{\rm post})$, keeping histories with the same final level distinct. For each trajectory, the fit uses geometric means from 24 equal logarithmic radius bins, with equal weight assigned to every occupied bin.

We compare fixed exponents by refitting their amplitudes on the same observations, with equal total weight per occupied radius bin. The smaller residuals for $0.629$ than for $2/3$ in all nine tested windows for each of the nine available histories favour $0.629$ over these intervals.

\begin{table}[ht]
\centering
\begin{tabular}{ccccc}
\toprule
$(\ell_{\rm pre},\ell_{\rm post})$ & $s$ & $\alpha_{\rm eff}$ & RMS ($0.629$) & RMS ($2/3$)\\
\midrule
(12,13) & $1.4472\pm0.0026$ & $0.6440\pm0.0011$ & 0.01772 & 0.02501\\
(12,14) & $1.4254\pm0.0038$ & $0.6351\pm0.0015$ & 0.00852 & 0.03381\\
(13,14) & $1.4320\pm0.0025$ & $0.6378\pm0.0010$ & 0.01083 & 0.03205\\
(13,15) & $1.4231\pm0.0021$ & $0.6342\pm0.0009$ & 0.00675 & 0.03615\\
(14,15) & $1.4063\pm0.0039$ & $0.6275\pm0.0015$ & 0.00749 & 0.04240\\
(14,16) & $1.4102\pm0.0036$ & $0.6290\pm0.0014$ & 0.00485 & 0.04049\\
(15,15) & $1.4031\pm0.0036$ & $0.6262\pm0.0014$ & 0.00835 & 0.04454\\
(15,16) & $1.4025\pm0.0052$ & $0.6260\pm0.0021$ & 0.00914 & 0.04957\\
(16,16) & $1.3998\pm0.0039$ & $0.6249\pm0.0015$ & 0.00897 & 0.04544\\
\bottomrule
\end{tabular}
\caption{Finite-window flux fits at $Oh=0.03$, reconstructed from the plotted observations. The reference window contains 11--25 observations and 11--24 occupied bins per trajectory. The quoted uncertainties are formal least-squares residual standard errors, calculated with $m-2$ degrees of freedom for $m$ occupied bins. The last two columns give the RMS residual in $\ln Q_j$ for the fixed exponents $\alpha=0.629$ and $\alpha=2/3$. For each fixed exponent, the amplitude is refitted to the same extracted observations, with equal total weight assigned to each occupied radius bin.}
\label{tab:flux}
\end{table}

Leaving the slope free gives $0.6249\le\alpha_{\rm eff}\le0.6290$ for the five histories with $\ell_{\rm pre}\ge14$ in the reference window, close to the value selected by the cone geometry. For $(\ell_{\rm pre},\ell_{\rm post})=(16,16)$, however, varying the fitting window gives $0.6118\le\alpha_{\rm eff}\le0.6304$ (Table~\ref{tab:windows}). The small formal errors are derived from residual scatter within individual fits and account for neither window dependence nor correlations along a trajectory.

\begin{table}[ht]
\centering
\small
\begin{tabular}{cccccc}
\toprule
$r_{j,\min}$ & $r_{j,\max}$ & $(12,13)$ & $(12,14)$ & $(13,14)$ & $(13,15)$\\
\midrule
0.0050 & 0.018000 & 0.6465 & 0.6389 & 0.6403 & 0.6362\\
0.0050 & 0.023952 & 0.6440 & 0.6351 & 0.6378 & 0.6342\\
0.0050 & 0.030000 & 0.6401 & 0.6298 & 0.6333 & 0.6301\\
0.0065 & 0.018000 & 0.6458 & 0.6356 & 0.6399 & 0.6365\\
0.0065 & 0.023952 & 0.6424 & 0.6306 & 0.6364 & 0.6333\\
0.0065 & 0.030000 & 0.6376 & 0.6241 & 0.6308 & 0.6280\\
0.0080 & 0.018000 & 0.6436 & 0.6336 & 0.6380 & 0.6356\\
0.0080 & 0.023952 & 0.6396 & 0.6285 & 0.6336 & 0.6312\\
0.0080 & 0.030000 & 0.6340 & 0.6217 & 0.6269 & 0.6246\\
\bottomrule
\end{tabular}
\par
\medskip
\begin{tabular}{ccccccc}
\toprule
$r_{j,\min}$ & $r_{j,\max}$ & $(14,15)$ & $(14,16)$ & $(15,15)$ & $(15,16)$ & $(16,16)$\\
\midrule
0.0050 & 0.018000 & 0.6319 & 0.6327 & 0.6313 & 0.6310 & 0.6304\\
0.0050 & 0.023952 & 0.6275 & 0.6290 & 0.6262 & 0.6260 & 0.6249\\
0.0050 & 0.030000 & 0.6229 & 0.6235 & 0.6217 & 0.6218 & 0.6197\\
0.0065 & 0.018000 & 0.6287 & 0.6305 & 0.6277 & 0.6277 & 0.6275\\
0.0065 & 0.023952 & 0.6241 & 0.6252 & 0.6215 & 0.6211 & 0.6209\\
0.0065 & 0.030000 & 0.6191 & 0.6180 & 0.6177 & 0.6173 & 0.6148\\
0.0080 & 0.018000 & 0.6250 & 0.6290 & 0.6248 & 0.6256 & 0.6254\\
0.0080 & 0.023952 & 0.6200 & 0.6232 & 0.6191 & 0.6185 & 0.6183\\
0.0080 & 0.030000 & 0.6148 & 0.6155 & 0.6140 & 0.6144 & 0.6118\\
\bottomrule
\end{tabular}
\caption{Effective exponents $\alpha_{\rm eff}$ obtained by varying the radius window, using the same 24-bin fitting procedure for each refinement history. Column headings give $(\ell_{\rm pre},\ell_{\rm post})$. Together, the two blocks contain the nine histories for which flux observations at $Oh=0.03$ are available in the figure.}
\label{tab:windows}
\end{table}

\section*{Radius at a fixed Reynolds number}

For the singular Worthington jet, a natural continuum cutoff may come from viscous stresses. However, the simulations do not resolve that cutoff length scale $r_j = r_m$, such that $Re_j \equiv v_jr_j/Oh \sim 1$. We compare the resolved jet across $Oh$ using its radius $r_j(Re_*)$ when the jet-base Reynolds number $Re_j$ reaches $Re_*=20$, 25 or 30. For a fixed cone angle, $v_jr_j\propto r_j^{(2\alpha-1)/\alpha}$ gives $r_j(Re_*)\propto(Re_*Oh)^{\alpha/(2\alpha-1)}$, so these choices alter the prefactor without changing the predicted power of $Oh$.

\begin{figure}[ht]
\centering
\includegraphics[width=\textwidth]{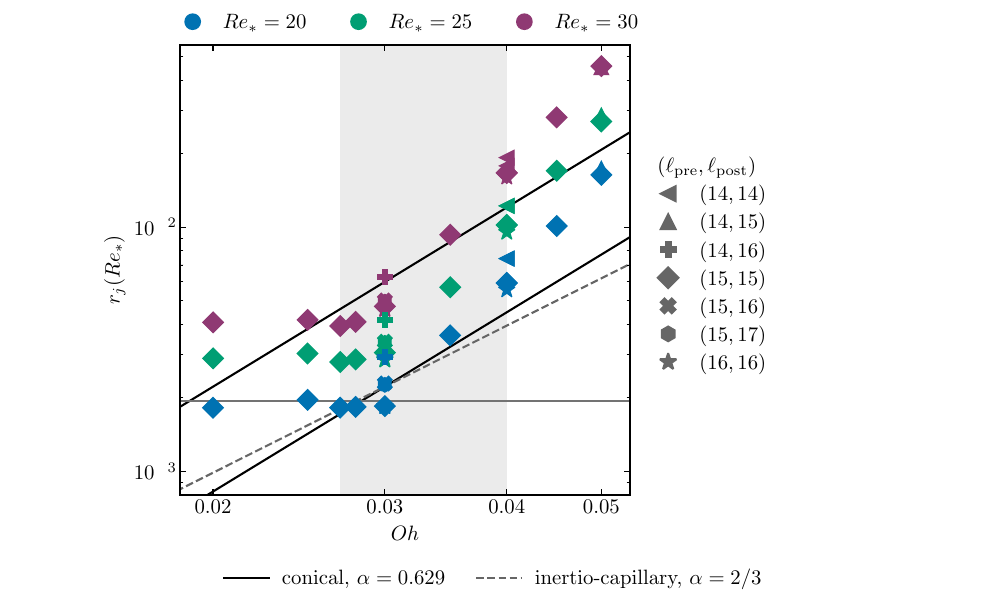}
\caption{Jet radius $r_j(Re_*)$ versus $Oh$ at $Re_*=20$ (blue), 25 (green) and 30 (purple), with symbols identifying the refinement pairs $(\ell_{\mathrm{pre}},\ell_{\mathrm{post}})$. Solid black lines show the conical scaling at $Re_*=20$ (lower) and 30 (upper), and the grey dashed line shows the inertio-capillary scaling. The horizontal grey line indicates the low-$Oh$ plateau, and the shaded band covers the conical regime $0.027<Oh<0.04$. The $Re_*=20$ points and guides are the same as in the inset of Fig.~2(b) and Fig.~5(c) of the Letter.}
\label{fig:threshold}
\end{figure}

We first take the median $Re_j$ within logarithmically spaced radius bins to suppress isolated transient spikes. We then interpolate in logarithmic coordinates between the two bin medians bracketing $Re_*$ to obtain $r_j(Re_*)$, using the same procedure for every $Oh$ and refinement history. For $Re_*=20$, the $(15,15)$ and $(16,16)$ histories give almost identical radii at $Oh=0.03$ and differ by about six percent at $Oh=0.04$ (Fig.~\ref{fig:threshold}). The corresponding radii at $Re_*=25$ and 30 also increase with $Oh$, while the differences between these two histories remain small. Neither the increase above the low-$Oh$ plateau nor the agreement between these histories therefore depends on choosing $Re_*=20$.

\section*{Interface collapse}

\begin{figure}[ht]
\centering
\includegraphics[width=\textwidth]{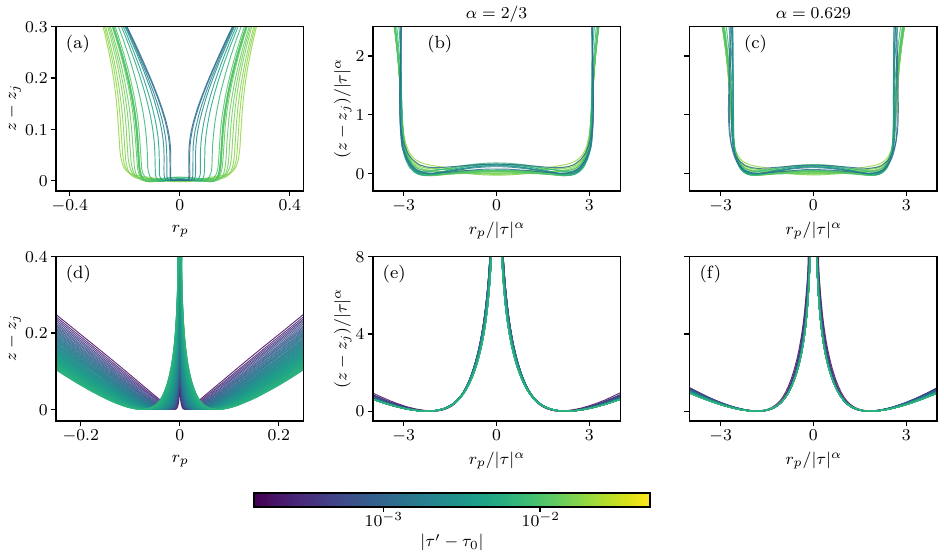}
\caption{Comparison of the two similarity exponents on the same interfaces. The upper row shows pre-inception cavity collapse; the lower row shows the post-inception jet. Columns show unscaled profiles, $\alpha=2/3$, and $\alpha=0.629$. In each rescaled panel, $\tau=\tau'-\tau_0$, with $\tau_0=0.494430$. Supplementary Videos~1 and~2 show the pre- and post-inception sequences.}
\label{fig:collapse-comparison}
\end{figure}

Figure~\ref{fig:collapse-comparison} compares the profiles shown in Fig.~3 of the Letter after rescaling with $\alpha=2/3$ and $\alpha=0.629$. Both exponents collapse the interfaces over the displayed pre- and post-inception windows, so the collapse of the shapes alone does not distinguish the underlying force balance.

\section*{Supplementary videos}
Video~1 shows the pre-inception cavity.\\
Video~2 shows the post-inception jet.\\
Video~3 provides a visual summary of the work.\\
Files: \texttt{SM1-pre-collapse.mp4}, \texttt{SM2-post-collapse.mp4}, and \texttt{SM3-summary.mp4}.